\documentclass[letterpaper,10 pt, conference]{IEEEtran}
\usepackage{amsmath,amssymb}
\usepackage{array} 
\usepackage[dvips]{graphicx}
\usepackage{xspace}
\usepackage{lscape}
\usepackage{here}
\usepackage{epsfig}
\usepackage{multirow}
\usepackage{color}
\usepackage{epstopdf}

\newcommand{\beq}{\begin{equation}}
\newcommand{\eeq}{\end{equation}}
\newcommand{\mx}[2]{\left[\begin{array}{#1}#2\end{array}\right]}
\newcommand{\barr}[2]{\begin{array}{#1}#2\end{array}}
\newcommand{\ds}{\displaystyle}
 \newcommand{\R}{\rm{I\kern-2pt R}}
\renewcommand{\theequation}{\arabic{equation}}
\newtheorem{thrm}{\bf Theorem}
\newtheorem{rmrk}{{\bf Remark}}
\newtheorem{ex}{\bf Example}
\newtheorem{crllry}{{\bf Corollary}}
\newtheorem{defi}{\bf Definition}
\newtheorem{prpstn}{{\bf Proposition}}
\newtheorem{lemme}{{\bf Lemma}}
\newenvironment{theorem}{\begin{thrm} \rm}{\end{thrm}}
\newenvironment{remark}{\begin{rmrk} \rm}{\end{rmrk}}

\newenvironment{definition}{\begin{defi} \rm}{\end{defi}}
\newenvironment{proposition}{\begin{prpstn} \rm}{\end{prpstn}}
\newenvironment{example}{\begin{ex} \rm}{\end{ex}}

\IEEEoverridecommandlockouts 

\begin{document}
\title{\vspace{4mm} Collision Free Navigation with Interacting, Non-Communicating Obstacles}

\author{Mario Santillo, Mrdjan Jankovic
\thanks{M. Santillo and M. Jankovic are with
  Ford Research and Advanced Engineering,
  2101 Village Road, Dearborn, MI 48124, USA,   {\tt\small e-mail: msantil3@ford.com, mjankov1@ford.com}}%
}

\maketitle

\begin{abstract}
\noindent
In this paper we consider the problem of navigation and motion control in an area densely populated with other agents. We propose an algorithm that, without explicit communication and based on the information it has,  computes the best control action for all the agents and implements its own. Notably, the host agent (the agent executing the algorithm) computes the differences between the other agents' computed and observed control actions and treats them as known disturbances that are fed back into a robust control barrier function (RCBF) based quadratic program. A feedback loop is created because the computed control action for another agent depends on the previously used disturbance estimate. In the case of two interacting agents, stability of the feedback loop is proven and a performance guarantee in terms of constraint adherence is established. This holds whether the other agent executes the same algorithm or not.  
\end{abstract}

\section{Introduction}
\noindent
We consider the problem of an autonomous vehicle or robot (referred to as the "host") operating in an environment with other autonomous or human-operated agents (referred to as "targets") that it has no communication with, yet interacts through {\em observed actions}. In contrast to the problem of avoiding stationary obstacles, the host's path depends on what other agents decide to do and this, in turn, depends on what the host agent does, and so on.  In many less structured traffic situations -- an open parking lot, a construction zone, a large roundabout -- both the path and timing of each participant has to be modified in response to what everyone else is doing. A similar situation could occur with mobile robots that share an open space with other  robots, humans, and/or human operated vehicles with no clear rules of precedence.
 
Most papers concerning collision avoidance with interacting agents are in the field of robotics. The Interacting Gaussian Process (IGP) method \cite{trautman} is a position-based approach that tackles the so-called "frozen robot problem" with humans acting as interacting moving obstacles to the mobile robot. The algorithm computes "likely" future positions of the host and the targets using a joint probability density function obtained from empirical data. No kinematic or dynamic model  was used -- it is just assumed that the robot and the pedestrians could occupy these precomputed positions at the scheduled time.
 
Another approach is based on reciprocal velocity obstacles \cite{snape,vandenberg} with each agent assuming half of the responsibility for collision avoidance. The agents need to reach an implicit agreement on what side to pass on. Otherwise, ``reciprocal dance" occurs where the agents switch sides being unable to decide.
If no permissible velocity can be found, the algorithm removes the farthest away robot or obstacle from consideration and repeats the calculation. Velocity is also used as the control variable in  Lyapunov-barrier function (LBF) based methods, such as \cite{mastellone, panagou}, where a non-holonomic unicycle robot model is considered with communication assumed in \cite{panagou}. The LBFs provide a guarantee of collision-free motion if their values are non-increasing. A side effect of LBFs being infinite at the boundary is that a region around an obstacle becomes inaccessible -- the algorithm sees obstacles as larger than they actually are. 
 
The acceleration-based approach with Control Barrier Functions (CBF) is considered in \cite{borrmann,wang}. The CBFs provide linear constraints on acceleration for a
Quadratic Program (QP). In the centralized version,
the controller knows everyone's desired acceleration and computes everyone's instantaneously-optimal
control action while adhering to the constraints. Compared to LBF's, the CBF with QP allows the distance to the obstacle boundary to decrease to 0, 
but  at a rate that decreases with the distance. An advantage over the reciprocal velocity approach is that the optimization is not bilateral (i.e. each pair of agents handled independently), but multilateral (all at once). A disadvantage is that it requires communication to all the agents. 
 In the decentralized, no-communication case also considered in \cite{borrmann,wang}, 
 the host computes only its own action assuming targets velocities are constant. Compared to the centralized controller 
 that has control inputs for all the agents at its disposal, 
the decentralized QP is more likely to be infeasible producing no solution. In such a situation, \cite{wang} proposes the 
(host) agent enter a separate ``braking mode" -- i.e. stop with full deceleration.
 
In this paper, we set up a quasi-centralized QP based on Robust Control Barrier Functions (RCBFs) (see \cite{jankovic_aut}) while still assuming no explicit communication between agents. 
Each agent computes optimal accelerations for all the agents, with information at its disposal, and implements its own control action. 
 Because the agents, in general, will not agree, the method considers
 other agent actions as disturbances for RCBF.  The disturbances could be assumed bounded or, as we have done in this paper, estimated on line
 as a difference between the actual target acceleration and the one computed by the host. 
 This creates a static loop (the disturbance is used to compute the target acceleration, while the acceleration is used to compute the disturbance) that has to be cut by 
inserting a unit delay, i.e. using the value from the previous sample, or a low pass filter creating internal controller states. Because the host predicts and then corrects the 
target acceleration, we refer to the algorithm as the Predictor-Corrector for Collision Avoidance (PCCA). Here are some
properties of the PCCA controller:
\begin{enumerate}
\item The QP for the PCCA controller has the same feasibility as the centralized controller of \cite{wang} because it has the control action  of all agents at its disposal, though only its own is actually applied. 
\item In the case of 2 agents, we prove that controller internal states are bounded and any error in constraint enforcement is of the order of the sampling time $\Delta t$.  Quantitatively similar performance guarantee  is  established even if the target agent is not interacting, while the host,  by running PCCA,  assumes it is. Based on results observed in simulations, we believe these properties might hold for multi-agent cases as well.
\item The PCCA computational complexity (for each agent as the host) is observed to be similar to the centralized controller. With $N_a$ denoting the number of agents, complexity theoretically scales as $N_a^4$ using the interior point method \cite{boyd}. 
\end{enumerate}
 
The paper is organized as follows. Section \ref{sec:rcbf} reviews the results related to QP with RCBF. Section \ref{sec:agents} introduces the model for the agents, equations describing the controller, and contains the main results. The simulations in Section \ref{sec:sims} consider cases of 2 agents, both running the PCCA algorithm or one running PCCA and the other non-interacting.\\

\noindent
{\bf Notation:} For a differentiable function $h(x)$ and a vector $f(x)$, 
$L_fh(x)$ denotes $\frac{\partial h}{\partial x} f(x)$.
A function $\alpha: {\R}^+ \rightarrow {\R}^+$ is of class $\cal{K}$ if it is continuous, zero at zero, and strictly increasing. 
A function $\gamma(t,\varepsilon)$ is said to be ${\mathcal O}(\varepsilon)$ if $|\gamma(t,\varepsilon)| \le \kappa |\varepsilon|$ for some $\kappa>0$ 
and for all sufficiently small $\varepsilon$.

\section{Robust Control Barrier Functions}\label{sec:rcbf}
\noindent
In this section, we review the concept of robust Control Barrier Functions introduced in \cite{jankovic_aut}. RCBFs are an extension of CBFs introduced in \cite{wieland} for systems with a bounded external disturbance $w(t) \in {\R^\nu}$, $ \| w(t)\| \le \bar w > 0$, of the form 
\beq \dot x= f(x) + g(x) u + p(x) w \label{dyn_w_dist} \eeq

One control objective for (\ref{dyn_w_dist}) is to regulate the 
state close to the origin (e.g. input-to-state stability (ISS)) 
and we assume that there is a known baseline controller 
$u_0$ that achieves this objective. The other control objective is to
keep the state of the system in an admissible set ${\mathcal C} = \{x: h(x)  \ge 0\}$. 
Below is the "zeroing" version of  Robust Control Barrier Functions (RCBF) definition. \\

\begin{definition} ({\em  Robust-CBF}) \ For the system (\ref{dyn_w_dist}), 
a differentiable function $h(x)$ is an RCBF with respect to the admissible set ${\mathcal C} = \{x: h(x)  \ge  0\}$ 
if there exists a class $\mathcal K$, Lipschitz continuous function $\alpha_h$ such that 
\beq L_gh(x) = 0 \ \Rightarrow  \ L_fh(x) - \|L_p h\| \bar w + \alpha_h(h(x)) > 0 \label{r-cbf} \eeq
\end{definition}

The definition states that, when the control input has no impact on
$\dot h \ (= L_fh+L_phw+L_ghu)$, under the worst-case disturbance,  the RCBF $h$ cannot decrease towards 0 faster than 
$\alpha_h(h)$. 

In the case of an unknown disturbance, we set up 
a barrier constraint for the worst case disturbance:
\beq F_1 = L_fh(x) - \|L_p h(x)\| \bar w + L_g h(x) u + \alpha_h(h(x)) \ge 0 \label{bar_w_const} \eeq
With a good estimate or a measurement of the disturbance $\hat w$, one can be less conservative by using
\beq F_2 = L_fh(x) + L_p h(x) \hat w + L_g h(x) u + \alpha_h(h(x)) \ge 0 \label{hat_w_const} \eeq
In both cases, the idea is to find a control $u$, that is 
as close as possible to the baseline control $u_0$ in the Euclidean
distance sense,  such that the 
barrier constraint is satisfied.  Compared to \cite{jankovic_aut}, the Lyapunov (CLF) constraint is removed and replaced
by $u_0$. While the Lyapunov constraint provides two tuning
parameters to adjust responsiveness of the controller (see \cite{jankovic_aut}), it would make the  analysis
carried out in the next section more difficult. From the theoretical standpoint, the change does not make a difference. \\ 

\noindent
{\bf Robust QP Problem}: Find the control $u$ 
 that satisfies
\beq \barr{l}{\ds \min_u \|u-u_0\|^2 \ \  {\rm subject \  to}  \\*[2mm]
 \ds F_i \ge 0, \ \  i = 1\  {\rm or}\  2 }\label{rQP} \eeq
where we select $F_1$ or $F_2$ (see (\ref{bar_w_const}) and (\ref{hat_w_const})) depending on whether we use an estimate/measurement
of the disturbance or the worst case upper bound. 
In either case, the following result applies. \\

\vspace{-0.1in}
 \begin{theorem} If   $h(x)$  is an RCBF for the system (\ref{dyn_w_dist})  then
\begin{enumerate}
\item	The Robust QP problem (\ref{rQP}) is feasible and the resulting control law  is Lipschitz continuous in  $\mathcal C$.
\item $\dot h(x) \ge - \alpha_h(h(x))$ for all $x\in {\mathcal C}$ and the set $\mathcal C$ is forward invariant. 
\item If the barrier constraint is inactive, $u = u_0$. 
As a result, if the barrier constraint is inactive for all
$t$ greater than some $t^*$ and $u_0$ is an ISS controller,
the closed loop system  is input to state stable with respect to the disturbance input $w$. 
 \end{enumerate}
\end{theorem}

The proof follows from Theorem 2 of \cite{jankovic_aut}, with appropriate modifications for 
zeroing RCBF and $u_0$ replacing the CLF constraint.\\

In contrast to the standard definitions of CBF (e.g. \cite{ames_pp, wieland}) and RCBF \cite{jankovic_aut},
the barrier function considered  for interacting agents later in this
paper has relative-degree two. That is, the control input and the disturbance input do not appear in $\dot h$, but in $\ddot h$ -- hence, relative degree two. Here, we briefly review the approach \cite{nguyen, xu_rel_deg}, which considers
\beq \ddot h + l_1 \dot h + l_0 h\ge 0 \label{r2_barrier}\eeq
as the QP constraint. The parameters
$l_0, l_1$ should be selected so that the two roots $\lambda_{1,2} = \frac{-l_1\pm\sqrt{l_1^2 -4l_0}}{2}$   of the polynomial
$s^2+l_1 s + l_0 = 0$ are negative real.
Then, if the barrier constraint (\ref{r2_barrier}) holds, it is not difficult to show that the set 
${\mathcal C}^* = \{ (x): h(x) \ge 0, h(x) \ge \frac{1}{\lambda_1 }\dot h(x) \}$, with
$\lambda_1$ being either one of the two roots, is forward invariant. Picking the smaller (more negative one) 
makes $\mathcal C^*$ larger.
With ${\mathcal C}^*\subset {\mathcal C}$ the original constraint $h(x) \ge 0$ will be satisfied.
 
Taking into account the disturbance, the RCBF condition for the system (\ref{dyn_w_dist}) with a barrier $h$ of
relative-degree two from $u$ and $w$ becomes 
\beq L_gL_f h = 0 \ \Rightarrow \ L_f^2 h - \|L_pL_f h\|\bar w +l_1 L_f h + l_0 h > 0 \label{r2-rcbf} \eeq
The barrier constraint that needs to be enforced for the admissible set to be forward invariant is
\beq F_1 = L_f^2 h - \|L_pL_fh \|\bar w + L_gL_f h u + l_1 L_fh + l_0 h \ge 0 \label{r2_const_barw} \eeq
in the case of unknown disturbance bounded by $\bar w$ or
\beq F_2 = L_f^2 h + L_fL_p h \hat w + L_gL_f h u + l_1 L_fh + l_0 h \ge 0 \label{r2_const_hatw} \eeq
in the case of known disturbance $\hat w$.

\section{Interacting Agents with No Communication}\label{sec:agents}
\noindent
In this section we develop the PCCA controller for agents that
cannot explicitly communicate, but still have to avoid
each other. Each of the $N_a$ agents is modeled as a circle of radius $r_0$ with the center motion
given by  the double integrator in two dimensions:
\beq \ddot X_i = u_{i}, \  i = 1,\ldots, N_a  \label{agent_i} \eeq
Here $X_i = [x_{i}, y_{i}]^T$ is agent $i$'s position in the plane and $u_i$ is its vector acceleration.
Each agent independently computes its preferred (base) control action $u_{0i}, i = 1,\ldots,N_a$, 
the acceleration  it would  implement if there were no obstacles.

The relative motion between any two agents $i$ and $j$ is given by
\beq \barr{l}{ \dot \xi_{ij}= v_{ij} \\ \dot v_{ij} = u_i - u_j} \label{diff_motion} \eeq
where $\xi_{ij} = X_i - X_j$ is the center-to-center vector displacement between the two agents and $v_{ij} = \dot X_{i} - \dot X_{j} $ is their relative velocity.
Our goal is to keep the $\|\xi_{ij}\|$ larger than $r \ge 2r_0$. When we use the 
distance $r$ strictly greater than $2 r_0$, we are providing a ``radius margin," counting on the resilience of the QP-CBF setup to
keep the actual constraint satisfied in the presence of uncertainties or disturbances (see \cite{xu} for details).
To this end, we define a barrier function
\beq h(\xi_{ij}) = \xi_{ij}^T\xi_{ij} - r^2  \label{h_ij} \eeq
with the goal to keep  it greater than 0. Because the barrier function $h$ has relative-degree two to all 4 inputs,
we apply the approach described in Section \ref{sec:rcbf} and form a barrier constraint:
\beq \ddot h + l_1 \dot h + l_0 h = a_{ij} + b_{ij}(u_i - u _j )\ge 0  \label{h_constr} \eeq
where $a_{ij} = 2v_{ij}^Tv_{ij}+ 2 l_1 \xi_{ij}^T v_{ij} +  l_0 (\xi_{ij} ^T\xi_{ij} - r^2)$, $b_{ij}  = 2 \xi_{ij} ^T$. 
This results in the reduced admissible set ${\mathcal C}^*_{ij}= \{ (\xi_{ij}, v_{ij}): h(\xi_{ij}) \ge 0, h(\xi_{ij}) \ge \frac{1}{\lambda_1 }\dot h(\xi_{ij}, v_{ij}) \}$ 
with $\lambda_1$ defined above. 

With the barrier constraints defined, a centralized controller, requiring full knowledge of $u_{0i}$'s, could be set up as follows:

\noindent
{\bf Centralized QP}: Find the controls $u_i, i=1,\ldots, N_a$ 
\beq \barr{l}{\ds \min_{u_{1}, \ldots u_{N_a} } \sum_{i=1}^{N_a} \|u_i -u_{0i}\|^2 \ \  {\rm subject \  to}  \\*[2mm]
 \ds a_{ij} + b_{ij}(u_i - u_j )\ge 0  \ \forall i,j =1,\ldots, N_a , i\not = j}\label{rQPc} \eeq
The QP solution could be computed by a central node and communicated to the agents, or each agent could solve 
it independently. If the QP problem is feasible, the control action would satisfy all the  barrier constraints (\ref{h_constr}) 
and guarantee collision-free operation  \cite{wang}. 
 
Without communication, the base control action $u_{0j}$ for the targets  are not available to the host  $i$ (i.e. the agent doing the
computation). In this case, each agent could implement an on-board decentralized controller:

\noindent
{\bf Decentralized QP} (for agent $i$): Find the control $u_i$ for the agent $i$
 that satisfies
\beq \barr{l}{\ds \min_{u_{i}}  \|u_i -u_{0i}\|^2 \ \  {\rm subject \  to}  \\*[2mm]
 \ds \chi a_{ij} + b_{ij}u_i\ge 0  \ \forall  j =1,\ldots, N_a , j\not = i}\label{rQPd} \eeq 
The value $\chi = 1$, as used in \cite{borrmann}, implies that the agent $i$ assumes full 
responsibility for avoiding all the targets (no reciprocal action assumed), while $\chi = 1/2$ assumes evenly shared responsibility. 
It was proven in \cite{wang} that, if all the agents execute the same reciprocal policy (e.g. $\chi = 1/2$ for agents with equal acceleration capability) and the QP is feasible 
for all of them, then collision avoidance is guaranteed. However, with only the host's control available to avoid all the $N_a-1$ targets,
either version of the decentralized policy might be infeasible, necessitating action external to the 
QP  (e.g. max braking proposed in \cite{wang}).

In this paper, we present a different QP policy (predictor-corrector for collision avoidance, PCCA) that is computed by each agent independently. It
takes into account everyone's constraints ($a_{ij}$'s and $b_{ij}$'s are known to all),  but
only the host's own base control -- the others are not known and 0's are used instead. Even with the known set of constraints,
the agents' computed actions need not agree and the constraint might still be violated. 
Instead of increasing the radius margin to avoid potential collisions,
we add a fictitious disturbance $\hat w$ to each target's acceleration
in (\ref{diff_motion}) and use the RCBF design of Section \ref{sec:rcbf}. It is easy to check that
$h_{ij}$'s are RCBFs because $b_{ij} \not = 0$ in $\mathcal C$, 
and, in fact, in any set that does not contain
$\xi_{ij}  = 0$ (that is, the circles don't completely overlap). Applying the RCBF design described in Section \ref{sec:rcbf}, 
we obtain:
 
\noindent
{\bf PCCA QP} (as computed by agent $i$): Find control actions $u_{ij}, j=1,\ldots, N_a$ 
such that
\beq \barr{l}{\ds \min_{u_{i1}, \ldots, u_{iN_a}} \left(  \|u_{ii} -u_{0i}\|^2 +\sum_{j=1, j\not = i}^{N_a} \|u_{ij}\|^2 \right)\  {\rm subject \  to}  \\*[2mm]
 \ds a_{jk} + b_{jk}(u_{ij}+ \hat w_{ij} - u_{ik} - \hat w_{ik})\ge 0  \  \\*[2mm]
 \ds \hspace*{1.3in} \forall  j,k =1,\ldots, N_a, j < k }\label{rQPp} \eeq
and implement its own: $u_i = u_{ii}^*$, where $u^*$ denotes the solution to (\ref{rQPp}).
  
\begin{figure}[htbp!]\vspace{-.05in}
    \centering
  \includegraphics[scale=0.52]{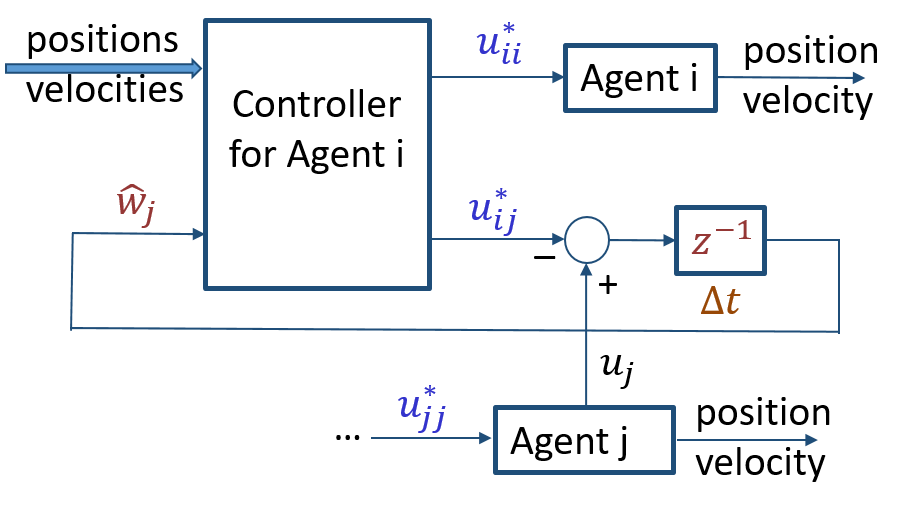}
    \vspace{-0.1in}
    \caption{Block diagram of the controller structure.}
    \label{fig:block_diag}
\end{figure}

As one could ascertain by comparing to Section \ref{sec:rcbf}, we have used the known-disturbance RCBF setup in (\ref{rQPp}). 
The disturbance estimate comes from 
comparing the control action for agent $j$ ($u^*_{ij} $) computed by the host (agent 
$i$) with the action agent $j$ actually implemented ($u_j$):
\beq \hat w_{ij} = u_j - u^*_{ij} \label {hatwj} \eeq 
where, obviously, $\hat w_{ii} \equiv 0$. Target acceleration could be obtained from an estimator (e.g. \cite{ansari})
or by approximate differentiation of the velocity signal. To compute $u_{ij}^*$ we need to know $\hat w_j$ and vice versa, creating a static (algebraic) feedback loop. 
We break the static loops by either using the value from the previous sample (the controller is implemented in discrete time) 
as illustrated in Fig. \ref{fig:block_diag} or
inserting a low-pass filter. In this paper we consider the former because it illuminates the stability mechanism more clearly. 
The latter would correspond to the  implementation where the target control action $u_j$ is obtained by approximate differentiation
of its velocity. 

Now we establish some properties for the PCCA QP controller. First, it is feasible if (and only if) the centralized QP is. 
This is because the set of constraints become identical if, for the former, we change variables $\bar u_{ij} = u_{ij} + \hat w_{ij}$.  Thus, if some $u_j^{c}, j= 1,\ldots, N_a$ is a feasible solution to (\ref{rQPc}) then $u_{ij}^{p} = u_j^{c} - \hat w_{ij}$
is a feasible solution to (\ref{rQPp}).

Next, we analyze the stability of the $\hat w$ feedback loops and establish a performance guarantee in terms
of constraint adherence.
Because of the complexity introduced by the QP solver, only the case with two interacting agents is analyzed below. 
The QP problem (\ref{rQPp}) for agent 1 is given by
\beq \barr{l}{\ds \min_{u_{11}, u_{12}} \|u_{11} - u_{01}\|^2 + \|u_{12}\|^2 \ \ {\rm such\  that} \\
             a + b u_{11} - b u_{12} - b \hat w_2 \ge 0 } \label{QP1} \eeq
where we have dropped the subscript "$ij$" from $a$, $b$, etc., because there is only one constraint.
This QP problem is feasible and can be solved in a closed form with the solution $[u_{11}^*, u_{12}^*]^T$ given by
\beq \mx{c}{ u_{11}^* \\ u_{12}^*}  = \mx{c}{ u_{01} \\ 0} + \mx{c}{ -\min\{0, \mu_1\} \\ \min\{0, \mu_1\}}  \frac{b^T}{2bb^T}  \label{u1star} \eeq
where $\ds \mu_1 = a + b u_{01}  - b\hat w_2$. Note that only $ u_{11}^*$ is implemented as $u_1$ to control the motion of agent 1, 
while $u_{12}^*$ is the estimate of what agent 2 is expected to do. 

The QP for agent 2 is set up the same way and the resulting
closed form solution $[u_{21}^*, u_{22}^*]^T$ is given by
\beq \mx{c}{ u_{21}^* \\ u_{22}^*} = \mx{c}{ 0 \\ u_{02} } + \mx{c}{ -\min\{0, \mu_2\} \\ \min\{0, \mu_2\}} \frac{b^T}{2bb^T} \label{u2star} \eeq
with $\ds \mu_2 = a - b u_{02}  +b\hat w_1$, $u_{22}^* $ used to control agent 2, and $u_{21}^*$ the agent 2 estimate of the 
expected action for agent 1. 

As discussed above,  the controller is implemented with a zero-order hold and a unit delay with sampling time $\Delta t$.
Thus, the estimates of the ``disturbances'' at time instant $t = k\Delta t$ are given by
\beq \barr{l}{ \hat w_1(k) = u_{11}^*(k-1)- u_{21}^*(k-1) \\*[2mm]
    		\hat w_2(k) = u_{22}^*(k-1) - u_{12}^*(k-1) } \label{wis} \eeq
With this setup we have the following result.

\begin{theorem}\label{thm2} The system (\ref{diff_motion}) with the controller (\ref{u1star}), (\ref{u2star}), (\ref{wis}) satisfies the following:
 \begin{enumerate}
 \item The control law computed by each agent is Lipschitz continuous in any set that does not include $\xi = 0$.
\item If $a$, $b$, $u_{01}$, and $u_{02}$ are bounded then $\hat w_1$ and $\hat w_2$ are bounded and, at time instants 
$t = k\Delta t$, satisfy
$\hat w_1(k)-\hat w_2(k)= u_{01} (k-1)- u_{02}(k-1) $.
 \item  If, in addition, the time derivatives of $u_{01}$ and $u_{02}$ are bounded, the closed loop system satisfies $h(k) \ge {\mathcal O} (\Delta t)$ and,
 if the sampling is  fast enough so that the change in $u_{01}$ and $u_{02}$ between two successive samples 
 is negligibly small, the admissible set is forward invariant.
\end{enumerate} 
\end{theorem}
  \ \\
\noindent
{\bf Proof}: \ {\bf Part 1}  --  Lipschitz continuity follows from the closed form solution and the fact that $bb^T = 4\xi^T  \xi \ge 16 r_0^2$.\\
{\bf Part 2} -- from (\ref{u1star}), (\ref{u2star}), (\ref{wis})
we have that
\beq \barr{l}{ \hat w_1(k)= \left . [u_{01}- (\min\{0, \mu_1\} - \min\{0, \mu_2\}  )  \frac{b^T}{2bb^T } ]\right |_{k-1} \\*[2mm]
\hat w_2 (k)= \left . [u_{02} + (\min\{0, \mu_2\} - \min\{0, \mu_1\} )  \frac{b^T}{2bb^T} ]\right |_{k-1} }\label{w_dyn} \eeq
The notation $(\cdot)|_{k-1}$ means that the values are taken at time $k-1$ (i.e. $t = (k-1)\Delta t$). 
Subtracting the two we obtain
\beq 
\hat w_1(k)-\hat w_2(k) = u_{01}(k-1) - u_{02}(k-1) \label{delta_w} \eeq
To prove boundedness of $\hat w_i$ we consider the
4 cases defined by signs of $\mu_i(k-1)$. \\
Case A ($\mu_1(k-1) > 0, \mu_2(k-1) > 0$): \ \ Directly from (\ref{u1star}), (\ref{u2star}) we obtain  $\hat w_i(k) = u_{0i}(k-1), \ i = 1,2$.  \\
Case B ($\mu_1(k-1) > 0, \mu_2(k-1) \le 0$): \ \ In this case we define $D =  \frac{b^T b}{2bb^T} $ and, using
$b w b^T = b^T b w$, obtain
\beq \hat w_1(k) = \left . \left[D \hat w_1 + (I-D)u_{01}+ \mu_{cc}\frac{b^T }{2bb^T} \right] \right |_{k-1} 
\label{w1_dyn} \eeq
where $\mu_{cc} = a + b u_{01} - b u_{02}$. The $\hat w_1$-dynamics in (\ref{w1_dyn}) is stable with Lyapunov function
$V = \hat w_1^T \hat w_1$ because $D^TD - I = \frac{1}{2} D - I \le -\frac{1}{2} I$ for all $b$.
Hence, $\hat w_1$ is bounded if the input signals are also bounded.
Boundedness of $\hat w_2$ follows from (\ref{delta_w}), but we shall write out the
dynamics for $\hat w_2$ needed later: 
\beq \barr{ll}{\hat w_2(k) = & \ds \left . \left[D \hat w_2 + (I-D)u_{02}+ \frac{a b^T }{2bb^T} \right] \right |_{k-1} + \\*[2mm]
& + \ D(k-1) (u_{01}(k-2)-u_{02}(k-2))  }
\label{w2_dyn} \eeq
Case C ($\mu_1(k-1) \le 0, \mu_2(k-1) > 0$): \ \ This case is symmetric to Case B and the same conclusion follows. \\
Case D ($\mu_1(k-1) \le 0, \mu_2(k-1) \le 0$): \ \  Define $\sigma_w = \hat w_1 + \hat w_2$ and note that 
\[ \barr{l}{ \hat w_1(k-1) = \frac{1}{2} [\sigma_w (k-1) + u_{01}(k-2) - u_{02}(k-2)] \\*[2mm]
    \hat w_2(k-1) = \frac{1}{2} [\sigma_w (k-1)  - u_{01}(k-2) + u_{02} (k-2)] } \]
 Using these two inequalities to replace $\hat w_i$'s with $\sigma_w$ in the $\mu_i \le 0$ inequalities that define 
 the Case D and rearanging terms we obtain
 \beq \! \! \barr{l}{ \left . (b\sigma_w)\right |_{k-1} \le \left. (2bu_{02}-2a )\right |_{k-1} \! - b(k-1) \left . (u_{01}-u_{02}) \right |_{k-2}\\*[2mm]
\left . (b\sigma_w)\right |_{k-1} \ge \left. (2a + 2bu_{01})\right |_{k-1}\! + b(k-1) \left . (u_{01}-u_{02}) \right |_{k-2} } \label{sigma_bound} \eeq
Note here that the right-hand sides are bounded by assumption. 
If we add the two equalities in (\ref{w_dyn}) we obtain
\[ \sigma_w(k) = \left . [2D \sigma_w + (I-2D) (u_{01} + u_{02}) ]\right |_{k-1}\]
Multiplying both sides by $b_1(k-1)$ and its orthogonal with the same magnitude $b_1^\perp(k-1)$  we obtain
\[ \barr{l}{ b(k-1) \sigma_w(k) = \left . (b \sigma_w) \right |_{k-1} \\*[2mm]
b^\perp(k-1) \sigma_w(k) = \left . [b^\perp (u_{01} + u_{02})] \right |_{k-1} }\]
Because the right-hand sides are bounded (the first from  (\ref{sigma_bound}), the second by assumption) 
and the matrix $\left . \mx{c}{b \\ b^\perp}\right |_{k-1}$ is invertible with determinant $\ge 4 r_0^2$, $\sigma_w$, the sum of $\hat w_i$'s,
is bounded. The difference $\hat w_2 - \hat w_1$ is also bounded by (\ref{delta_w}) and, hence,
$\hat w_i$'s are bounded  in Case D.

Thus far, we have shown that $\hat w_i(k)$'s are bounded if they are computed in Cases A and D. 
They are also bounded if they stay in Cases B and C due to their input-to-state stable dynamics. 
Finally, we consider the possibility of switching between Cases B and C. 
By considering (\ref{w1_dyn}), (\ref{w2_dyn}) and using the symmetry, we obtain that in both cases the dynamics are
\[ \hat w_i(k) = \left . (D\hat w_i)\right |_{k-1} + \chi_{i,{\rm B/ C}}(a,b,u_{01}, u_{02}), \  i = 1,2 \]
Hence, since the state matrix $D$ is the same in all cases, we can use the same time-invariant Lyapunov function $V$ defined above, and the $\hat w_i$ dynamics are input-to-state stable. The switching between cases B and C only changes $\chi_{i,{\rm B/ C}}$ which are bounded by assumption. \\
{\bf Part 3} --  
Recall that the actual barrier constraint that would guarantee forward invariance of $\mathcal C^*$ for (\ref{diff_motion}) is
$\mu_{cc} = a+bu_1-bu_2 \ge 0$ and consider again the 4 cases:\\
Case A ($\mu_1(k) > 0, \mu_2(k) > 0$): \ \  The control applied is $u_1 = u_{01}$ and
$u_2 = u_{02}$. Adding $\mu_1>0$ and $\mu_2>0$ and using (\ref{delta_w}),
we obtain $a+ bu_{01} - bu_{02} > -b\Delta_{u_0} = {\mathcal O} (\Delta t)$, where $\Delta_{u_0}(k) = \left . (u_{01} - u_{02}) ]\right |_{k} -
\left . (u_{01} - u_{02}) ]\right |_{k-1}$ is the change in $u_{01} - u_{02}$ between the current and previous samples.\\
Case B ($\mu_1(k) > 0, \mu_2 (k)\le 0$): \ \ In this case  the actual barrier constraint  is
$a+bu_1 - bu_2 = \frac{1}{2} a + b u_{01} - \frac{b}{2} u_{02} - \frac{b}{2} \hat w_1 = \frac{1}{2} (a -b\hat w_2 + b u_{01} ) + b\Delta_{u_0}
= \frac{1}{2} \mu_1+ b\Delta_{u_0}
> b\Delta_{u_0} = {\mathcal O}(\Delta t)$
where we have used $\hat w_1-\hat w_2 = u_{01} - u_{02} - \Delta_{u_0}$. \\
Case C is analogous to Case B. \\
Case D ($\mu_1 \le 0, \mu_2 \le 0$): \ \ The actual barrier constraint is $a + bu_1 - bu_2 = a + bu_{11}^* - bu_{22}^*= \frac{b}{2}(u_{01}-
u_{02}) - \frac{b}{2}(\hat w_1 - \hat w_2) = b\Delta_{u_0} = {\mathcal O}(\Delta t)$.

In all four cases we have that $a + bu_1 - bu_2  = \ddot h + l_1 \dot h + l_0 h \ge {\mathcal O} (\Delta t)$ 
and the barrier function satisfies $h(t)\ge  {\mathcal O} (\Delta t)$.
If $\| \Delta_{u_0} \|$ is negligibly small, then the admissible set is forward invariant. 
This completes the proof. 
\hspace*{\fill} $\bigtriangledown$ \\

\begin{remark} The proof establishes that in all four cases $\ddot h + l_1 \dot h + l_0 h \ge b\Delta_{u_0}$.
With $\|b\| \le \beta$ and $\|\dot u_{0i}\| \le M,\  i = 1,2$ for some $\beta, M > 0$ (as assumed in Theorem 2), we have that 
$\ddot h + l_1 \dot h + l_0 h \ge -2 \beta M \Delta t$ providing a way to assess how small the sampling time needs to be,
or, for a given sampling time, compute the necessary radius margin. 
\end{remark} 
\ \\
The above result tells us what happens when the agents are cooperating. Next, we analyze the 
PCCA controller when the situation is the opposite -- agent 2 is not interacting, while agent 1 assumes it is.
That is, the control action of agent 1 is computed by (\ref{u1star}) while agent 2 simply
applies its unconstrained control action $u_2 = u_{02}$. Note that only $\hat w_2$ is computed in this case.

\begin{proposition}\label{prop1} The system (\ref{diff_motion}), with the controller $u_1$ computed by the
controller (\ref{u1star}) and $u_2 = u_{02}$ assumed bounded and differentiable, satisfies the following:
for any $T>t_0$ there exists  $\Delta t^*$ such that for all sampling times  $\Delta t < \Delta t^*$
\beq h(\xi(t)) \ge {\mathcal O}(\Delta t), \ \forall t \in [t_0, T] \label{h_singpert} \eeq
\end{proposition}

\noindent {\bf Proof:} \ \ 
The proof relies on the singular-perturbation approach (see, for example, Chapter 11 in \cite{khalil}). 
To this end, we introduce a singular perturbation parameter $\varepsilon$: $\Delta t = \varepsilon \Delta t_0$. 
To simplify notation, we assume $\Delta t_0 = 1s$. Using (\ref{w1_dyn}), the symmetry between Cases B and C, and rearranging terms, we obtain the fast dynamics of the singularly perturbed system:
\beq \varepsilon \dot {\hat w}_2 = A \hat w_2 - A u_{02} + \min\{0, \mu_{cc}\}\frac{b^T}{2bb^T} + {\mathcal O}
(\varepsilon) \label{w2_fast} \eeq
where $A = D-I$ with $D$ and $\mu_{cc}$  defined  for (\ref{w1_dyn}) in the proof of Theorem 2. 
The fast $\hat w_2$ dynamics is 
exponentially stable uniformly in the  "slow" variables $a$, $b$, and $u_{0i}$. 
The slow dynamics is given by (\ref{diff_motion}) with 
\[ u_1 - u_2 = u_{01} - u_{02} -  \min\{0, a+bu_{01} -b\hat w_2\}\frac{b^T}{2bb^T} + {\mathcal O}
(\varepsilon) \]

Using the standard singular-perturbation approach, we assume that the slow variables are frozen 
and compute the quasi steady-state behavior of $\hat w_2$ at $\varepsilon = 0$. For the slow dynamics $(\xi, v)$,
we are only interested in $b \hat w_2$. Solving (\ref{w2_fast}) for the equilibrium of $\hat w_2$ at 
$\varepsilon = 0$, denoted by $\bar w_2$, and
using $b(I-D) = \frac{b}{2}$, we have
\beq b \bar  w_2 = \left\{ \barr{ll}{ bu_{02} \ & {\rm if} \ \mu_{cc} > 0 \\
  			-a - b u_{01} + 2 b u_{02}   & {\rm if} \ \mu_{cc}  \le 0} \right. \label{bar_z} \eeq
Using the quasi steady-state value $b \bar w_2$ in the reduced model, we obtain
\beq \dot {\bar v} = \left\{ \barr{ll}{ u_{01} - u_{02} \ & {\rm if} \ \mu_{cc}> 0 \\
  			u_{01} - u_{02} - \mu_{cc} \frac{b^T}{bb^T}  & {\rm if} \ \mu_{cc}  \le 0} \right. \label{bar_bv} \eeq
With $\dot {\bar \xi} = \bar v$, the reduced $(\bar \xi, \bar v)$ dynamics
are identical to the dynamics that the centralized QP controller would have produced. 
It is easy to check that they satisfy the barrier constraint (\ref{h_constr}) and, hence, achieve 
$h(\bar \xi) \ge 0$. 

The singular-perturbation result (Theorem 11.1 in \cite{khalil}) states that, on any finite time interval $t\in[t_0, T]$, for sufficiently 
small $\varepsilon$, the difference between the solution $\xi(t,\varepsilon)$ of the original
system and the solution $\bar \xi(t)$ of the reduced system is $ {\mathcal O}(\varepsilon)$
uniformly in $t \in [t_0, T]$. As a result,
\beq h(\xi(t,\varepsilon)) \ge {\mathcal O}(\varepsilon) \label{h_perturbed} \eeq
Subsituting $\Delta t = \varepsilon \Delta t_0 = \varepsilon$ completes the proof of the Proposition. 
\hspace*{\fill} $\bigtriangledown$ \\

Because $ {\mathcal O}(\varepsilon)$ is not sign definite, with one agent non-interacting, 
the PCCA controller requires a radius margin of the order $\Delta t$ to ensure collision-free operation. 
The results of item 3 in Theorem \ref{thm2} and the singular-perturbation result of Proposition \ref{prop1}
look qualitatively very similar, but are quantitatively
different.  With both agents cooperating, (\ref{delta_w}) provides that, for constraint adherence,  
the control adjusts to changes with one sample lag.
 With one agent non-interacting, the other agent takes only a part of the responsibility in each step. Becasue
 the other agent is not reacting, it eventually takes full responsibility with the time constant of $2\Delta t$. 
 We will see the difference in simulations in the next section where the former case does not require 
 any radius margin, while the latter does.

\section{Simulation Results}\label{sec:sims}
\noindent
We illustrate the results by simulation in the cases of two agents maneuvering in an enclosed area. In each case, the radius of the agents' circles is taken to be $r_0 = 2$, and hence, including the radius margin, $r\ge 4$ is used for computation of the barrier constraint $h$. The controller sample time is $\Delta t = 50$ ms. 
For computation of the QP constraints \eqref{h_constr}, we choose $l_0 = 6$ and $l_1 = 5$ to satisfy $l_1^2 \ge 4l_0$. The baseline controller $u_{0i}$ for each agent is computed by the Linear Quadratic Regulator (LQR) with $Q=4I_4$ and $R=I_2$ intended to bring the agent to a preassigned destination.

\begin{example}[Two Interacting Agents]\label{ex1}
In this example, we illustrate the controller operation by considering the completely symmetric case of two agents approaching each other head-on while heading towards each other's initial position.  Because the PCCA controller is continuous and the case completely symmetric, the agents brake instead of steer and stop without colliding (second plot from the top in Fig. \ref{fig:Ex1_h}). Stimulated by numerical rounding error or, more likely, precision setting of the solver, at some point (third plot in Fig. \ref{fig:Ex1_h})) they start moving again and pass one another without collision. Note that the agents only run the PCCA algorithm, with no preference for the passing side and no external de-conflicting mechanism such as in \cite{celi} or \cite{wang}. 

%
\begin{figure}[tp!]\vspace{-.2in}
    \centering
    \includegraphics[scale=0.93]{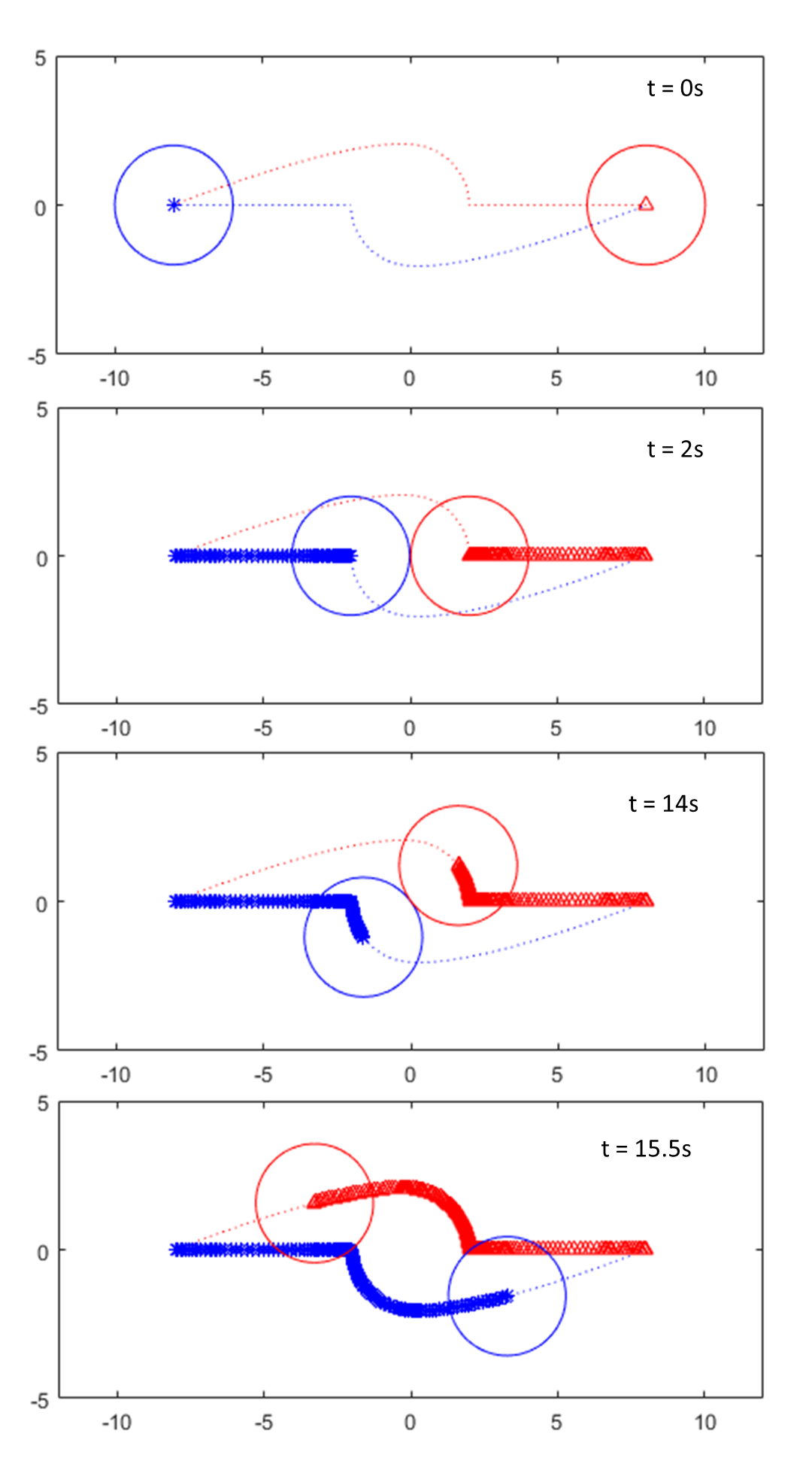}
    \vspace{-0.25in}
    \caption{Example \ref{ex1}: A series of snapshots in time for two interacting agents executing PCCA controllers in the completely symmetric head-on approach. }
    \label{fig:Ex1_h}
\end{figure}
%
 
\end{example}

\begin{example}[Two Agents as Pursuer/Evader]\label{ex2}
In this example agent 1 (evader) acts as before, that is, executes the controller \eqref{u1star}, but agent 2's (pursuer) goal is modified to intercept agent 1. 
That is, agent 2 operates without 
any collision avoidance action and its only control input  is the LQR control $u_{02}$ with the destination being the current location of agent 1.

Figure \ref{fig:Ex2} displays time snapshots of the pursuer/evader scenario. We see that, since agent 2 continually pursues agent 1, it is impossible for agent 1 to reach its destination and come to a rest. 
%
\begin{figure}[htbp!]
    \centering
    \includegraphics[trim={0 0.15in 0 0},clip,scale=0.7]{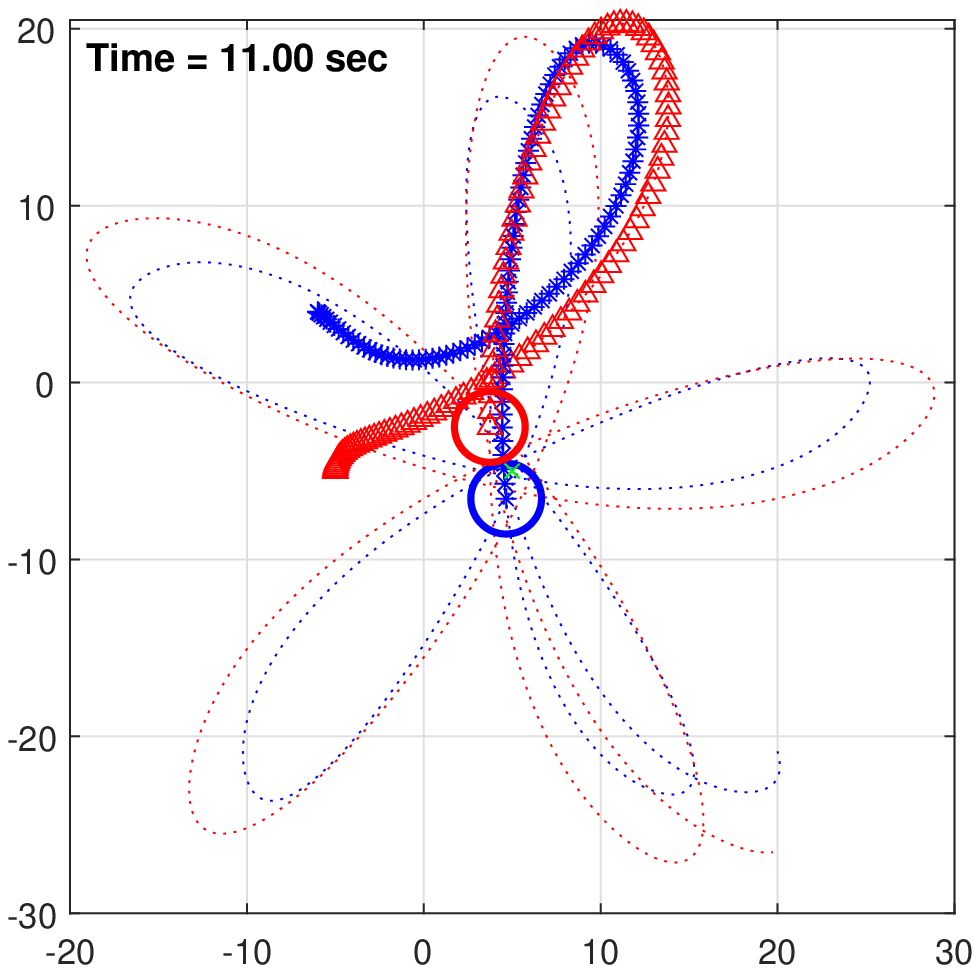}
    \includegraphics[trim={0 0 0 0},clip,scale=0.7]{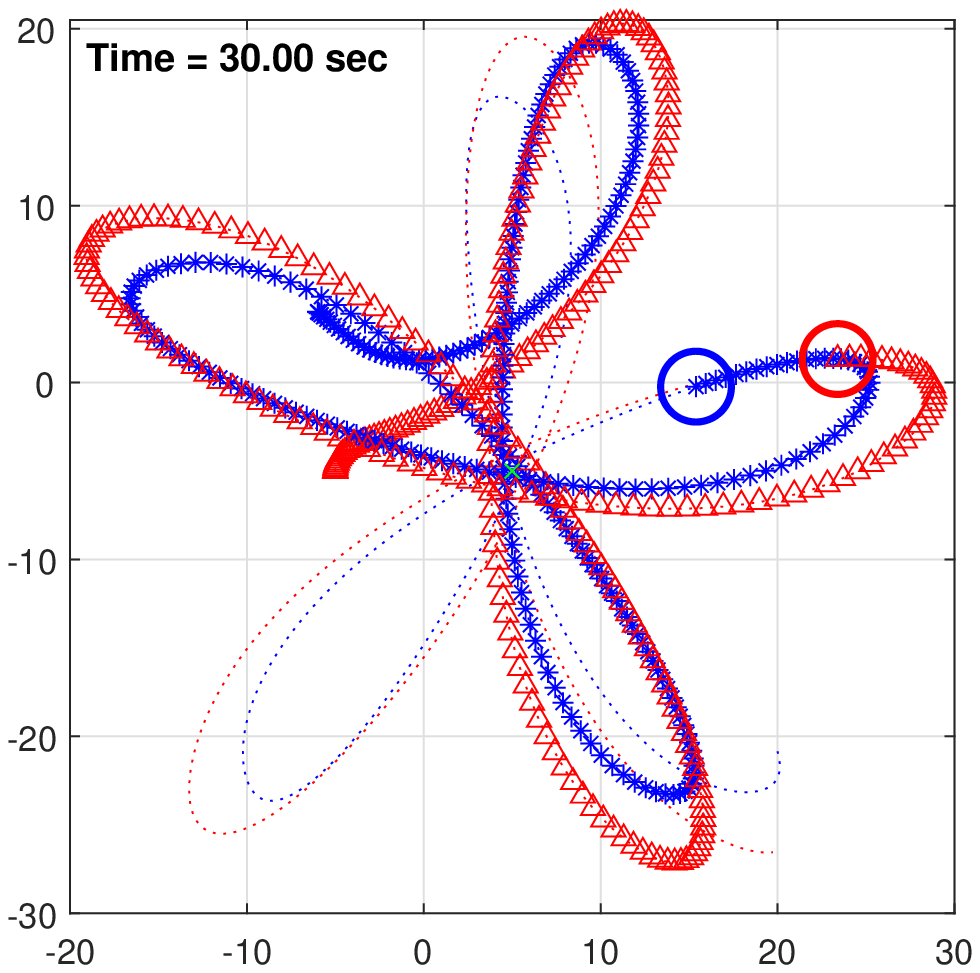}
    \caption{Example \ref{ex2}: Pursuer/Evader scenario does not allow agent 1 (evader) to achieve its destination while agent 2 (pursuer) follows.}
    \label{fig:Ex2}\vspace{-0.0in}
\end{figure}
%
With only one agent interacting, the analysis at the end of Section \ref{sec:agents} suggests that we must add a margin ($r > 2r_0$) to ensure collision-free operation. In this case, we add a margin equivalent to just over 1\% of the agent's radius such that the barrier for actual physical collision avoidance, $h_{r0}(x) = \xi^T\xi - 2r_0^2$, remains positive. This is shown in Figure \ref{fig:Ex2_h}, where the barrier $h_{r0} \ge 0$. Additionally, as the simulation sample time reduces by a factor of 5 (50 ms to 10 ms), the radius margin necessary to ensure collision-free operation reduces by a factor of 6, confirming the results presented in Section \ref{sec:agents}.

\begin{figure}[htbp!]\vspace{-0.1in}
    \centering
    \includegraphics[scale=0.65]{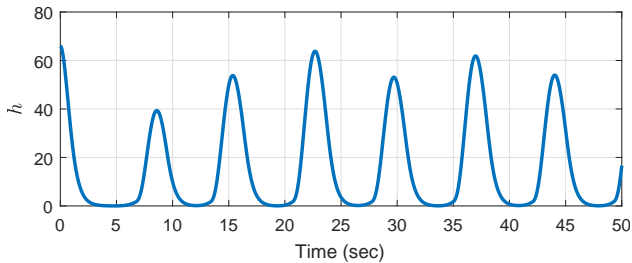}
    \vspace{-0.1in}
    \caption{Example \ref{ex2}: Barrier $h_{r0} \ge 0$ indicates no collision between the two agents with a 1\% radius margin added.}
    \label{fig:Ex2_h}\vspace{-0.1in}
\end{figure}
\end{example}

\section{Conclusions}
\noindent
We considered the problem of navigation and motion control in an area shared with other agents. Without explicit communication, the control algorithm utilizes observed acceleration of each target agent and compares it to its ``best" action as computed by the host. The difference between observed and computed accelerations guides a modification to the action taken by the host and a recomputed best action for the targets. The algorithm is shown to be stable and to avoid collisions if the sampling is fast enough. The result applies in the case
the other agent is cooperating, as well as when it is not cooperating as assumed. The design allows a 
tight representation of obstacles which is beneficial in densely populated  operating areas.


\end{document}